\documentclass[letter,prd,aps,twocolumn,floatfix,superscriptaddress]{revtex4}

\usepackage{amssymb,amsmath,graphicx}
\usepackage{hyperref}
\usepackage[usenames,dvipsnames]{color}
\usepackage{array}
\makeatletter
\renewcommand*{\p@section}{\S\,}
\renewcommand*{\p@subsection}{\S\,}

\makeatother

\def\jnl@style{\it}
\def\aaref@jnl#1{{\jnl@style#1}}

\def\aaref@jnl#1{{\jnl@style#1}}

\def\aj{\aaref@jnl{AJ}}                   
\def\araa{\aaref@jnl{ARA\&A}}             
\def\apj{\aaref@jnl{ApJ}}                 
\def\apjl{\aaref@jnl{ApJ}}                
\def\apjs{\aaref@jnl{ApJS}}               
\def\ao{\aaref@jnl{Appl.~Opt.}}           
\def\apss{\aaref@jnl{Ap\&SS}}             
\def\aap{\aaref@jnl{A\&A}}                
\def\aapr{\aaref@jnl{A\&A~Rev.}}          
\def\aaps{\aaref@jnl{A\&AS}}              
\def\azh{\aaref@jnl{AZh}}                 
\def\baas{\aaref@jnl{BAAS}}               
\def\jrasc{\aaref@jnl{JRASC}}             
\def\memras{\aaref@jnl{MmRAS}}            
\def\mnras{\aaref@jnl{MNRAS}}             
\def\pra{\aaref@jnl{Phys.~Rev.~A}}        
\def\prb{\aaref@jnl{Phys.~Rev.~B}}        
\def\prc{\aaref@jnl{Phys.~Rev.~C}}        
\def\prd{\aaref@jnl{Phys.~Rev.~D}}        
\def\pre{\aaref@jnl{Phys.~Rev.~E}}        
\def\prl{\aaref@jnl{Phys.~Rev.~Lett.}}    
\def\pasp{\aaref@jnl{PASP}}               
\def\pasj{\aaref@jnl{PASJ}}               
\def\qjras{\aaref@jnl{QJRAS}}             
\def\skytel{\aaref@jnl{S\&T}}             
\def\solphys{\aaref@jnl{Sol.~Phys.}}      
\def\sovast{\aaref@jnl{Soviet~Ast.}}      
\def\ssr{\aaref@jnl{Space~Sci.~Rev.}}     
\def\zap{\aaref@jnl{ZAp}}                 
\def\nat{\aaref@jnl{Nature}}              
\def\iaucirc{\aaref@jnl{IAU~Circ.}}       
\def\aplett{\aaref@jnl{Astrophys.~Lett.}} 
\def\apspr{\aaref@jnl{Astrophys.~Space~Phys.~Res.}}
\def\bain{\aaref@jnl{Bull.~Astron.~Inst.~Netherlands}} 
\def\fcp{\aaref@jnl{Fund.~Cosmic~Phys.}}  
\def\gca{\aaref@jnl{Geochim.~Cosmochim.~Acta}}   
\def\grl{\aaref@jnl{Geophys.~Res.~Lett.}} 
\def\jcp{\aaref@jnl{J.~Chem.~Phys.}}      
\def\jgr{\aaref@jnl{J.~Geophys.~Res.}}    
\def\jqsrt{\aaref@jnl{J.~Quant.~Spec.~Radiat.~Transf.}}
\def\memsai{\aaref@jnl{Mem.~Soc.~Astron.~Italiana}}
\def\nphysa{\aaref@jnl{Nucl.~Phys.~A}}   
\def\physrep{\aaref@jnl{Phys.~Rep.}}   
\def\physscr{\aaref@jnl{Phys.~Scr}}   
\def\planss{\aaref@jnl{Planet.~Space~Sci.}}   
\def\procspie{\aaref@jnl{Proc.~SPIE}}   

\newcommand{\Alfven}{Alfv\`en}

\def\Alfven{Alfv{\'e}n\, }

\newcommand{\had}{{\sc had}}

\begin{document}

\date{\today}
\title{Electromagnetic Luminosity of the Coalescence of Charged Black Hole Binaries}

\author{Steven L. Liebling}
\affiliation{Department of Physics, Long Island University, Brookville, New York 11548, USA}
\author{Carlos Palenzuela}
\affiliation{Departament de F\'isica, Universitat de les Illes Balears and Institut d'Estudis Espacials e Catalunya, Palma de Mallorca, Baleares E-07122, Spain}


\begin{abstract}
The observation of a possible electromagnetic counterpart by the Fermi GBM group
to the aLIGO detection of the merger of a black hole binary has spawned a number
of ideas about its source. 
Furthermore, observations of fast radio bursts~(FRBs) have similarly resulted in a range
of new models that might endow black hole binaries with electromagnetic signatures.
In this context, even the unlikely idea that astrophysical
black holes may have significant charge is worth exploring, and here we present results
from the simulation of weakly charged black holes as they orbit and merge. 
Our simulations suggest that a black hole binary with mass comparable to that observed
in GW150914 could produce the level of electromagnetic luminosity observed by Fermi\,GBM ($10^{49}$\,ergs/s) with a
non-dimensional charge of $q \equiv Q/M = 10^{-4}$ assuming good radiative efficiency.
However even a charge such as this is difficult to imagine avoiding neutralization long enough for the binary
to produce its electromagnetic counterpart, and so this value would likely serve simply as an upper bound.
On the other hand, one can equivalently consider the black holes as having
acquired a magnetic monopole charge that would be easy to maintain and would generate an identical
electromagnetic signature as the electric charges. The observation of such a binary would
have significant cosmological implications, not the least of which would be an explanation
for the quantization of charge itself. We also study such a magnetically charged binary in the force-free regime
and find it much more radiative, reducing even further the requirements 
to produce
the counterpart.
\end{abstract}

\maketitle


\section{introduction}

The exciting first direct detection of gravitational waves occurred recently~\cite{Abbott:2016blz}, beginning
the era of gravitational wave astronomy. 
Surprisingly, aLIGO's first detected event was not a binary neutron star merger as expected, but instead the merger of two quite massive
black holes.
Adding to the intrigue
was a follow-up detection by the Fermi~GBM team
in the hard X-ray band
that appears to be a coincident detection~\cite{Connaughton:2016umz}.
If indeed these X-rays came from aLIGO's
stellar-mass (albeit large) binary black hole~(BBH), new and possibly radical ideas are needed to explain
how a BBH such as this would have produced such an electromagnetic signal.

The prospect that the first direct GW detection also resulted in a coincident detection is a great harbinger
of what is to come from multi-messenger astronomy. However, significant skepticism surrounds this claim~\cite{Greiner:2016dsk}.
Fermi~GBM faced  a couple of obstacles for their follow-up: (i)~The aLIGO trigger included a huge swath of sky (roughly
$600$~square degrees), and (ii)~the detector (with a field of view of roughly 70\% of the sky) only observed roughly 75\% of the GW150914 skymap.
The signal could meet the threshold for claiming a detection only in combination with
aLIGO's detection.
In particular, they find a signal $0.4$\,s after the peak gravitational wave signal lasting for $1.0$\,s.
Two other detectors found no detection, including 
AGILE~\cite{Tavani:2016jrd} and
INTEGRAL~\cite{Savchenko:2016kiv}.
We leave it to others to argue one way or another about coincidence, and propose here simply that
possibilities such as this are worth exploring if only because future surprises are sure to continue.

In response, a number of scenarios in which a BBH could produce such an electromagnetic signature have been proposed. 
These scenarios include: the collapse of a massive star into two clumps which subsequently
collapse to black holes~\cite{Loeb:2016fzn}, the formation of a ``dead'' disk about one of the
black holes that powers a burst during the pre-merger phase~\cite{Perna:2016jqh}, the merger
of two very massive, low metallicity stars~\cite{Woosley:2016nnw}, and the relativistic outflow of material
into an ambient medium~\cite{Yamazaki:2016fyr}.
A critical examination of aspects of some of these models is presented in Ref.~\cite{Lyutikov:2016mgv}.

A proposal by Zhang~\cite{Zhang:2016rli}  (see also a related proposal by Ref.~\cite{Fraschetti:2016bpm})
argues that a charged BBH  could have produced the signal.
The motion of a charged black hole constitutes a current loop that drives the resulting Poynting flux.
This flux interacts with the surrounding environment which introduces a time lag corresponding 
to the $0.4$\,s observed by Fermi\,GBM.
Although the charge required to produce the observed energy flux
is small in non-dimensional units, the charge in absolute terms is still $5\times 10^{16}$\,C.
In any case, one does not expect to find astronomical objects with a net charge for long times
because of the huge electromagnetic forces trying to neutralize such objects. 

Regardless of the status of this Fermi~GBM detection, the observations of fast radio
bursts~(FRBs) similarly require new ideas about possible electromagnetic signals. Such
bursts are extra-galactic, radio transients with durations of just a few milliseconds~\cite{Petroff:2016tcr}.
Although one reported FRB repeats~\cite{Spitler:2016dmz,Scholz:2016rpt} and hence cannot be explained by cataclysmic events,
a sub-class of FRBs may still be explainable by binary mergers~\cite{Petroff:2016tcr} (also see Ref.~\cite{Callister:2016vtl}).
For example, Ref.~\cite{Liu:2016olx} proposes the discharging of charged black holes as a source of FRBs.

And although a proposal such as this 
naturally merits much skepticism,  we study this scenario both because the possible coincident
detection calls for unexpected ideas and because the dynamics of charged black holes
are interesting in their own right (having been studied in the head-on collision case, for example, by
Refs.~\cite{Zilhao:2012gp,Zilhao:2013nda,Zilhao:2014wqa}). 
Indeed, it makes sense to explore a variety of implications
of this detection if only to set bounds on black hole charge~\cite{Yunes:2016jcc,Giudice:2016zpa}.

Another motivation arises from studies of the interaction of
black holes with an external magnetic field that showed a significant Poynting
flux powered by the kinetic energy of the black holes. This coupling is the generalization of
the Blandford-Znajek effect for rotating black holes and it is quite a robust effect~\cite{Palenzuela:2010nf,Neilsen:2010ax}. However, because black holes
generally do not support their own magnetic field, such studies usually assume an external source of the
field such as a circumbinary disk. However, a black hole can support a monopolar 
electric or magnetic charge, and so this work is a natural extension of such studies
(and these studies also motivate the magnetic reconnection model of Ref.~\cite{Fraschetti:2016bpm}
 within a charged BBH scenario).

Despite the natural skepticism about astrophysical charged black holes, 
a number of proposals 
explain how a black hole might maintain a charge.
Ref.~\cite{25354} (in particular see Ch. 11) describes that certain
stable structures can shield a BH from electric discharge, such as a ring
of charged particles (i.e., with charge opposite that of the black hole) rotating at a certain radius.
Another argument begins with a BH rotating with angular momentum $J$ immersed in an external magnetic
field of strength $B_0$.  Such a BH will become charged up to $Q=2B_0 J$
so that with the Kerr limit, $Q/M \le 2B_0 M$~\cite{Wald:1974np}.
If somehow a black hole binary found itself within a very strong 
magnetic field supported by a circumbinary disk, the charge could similarly
be significant.
There is also the possibility that a star becomes charged prior to collapse to a black hole and it maintains the charge during the collapse (see  Ref.~\cite{Khuri:2015xpa} and references therein).
It should be said that these scenarios generally apply for extremely small charges.

Some models of dark matter incorporate a dark electromagnetism that could
potentially result in charged BHs~\cite{Ackerman:mha,Cardoso:2016olt}.
To be relevant to the Fermi\,GBM detection, the dark Poynting flux would somehow
have to convert into hard (non-dark) X-rays.

Finally, black holes may accrete cosmological magnetic monopoles~\cite{Stojkovic:2004hz,Preskill:1984gd}. In the region
outside such magnetically charged black holes, the symmetry between electric and magnetic fields
in the absence of charges and currents (outside the event horizons),
would imply that their dynamics would be identical to a charged BBH binary.
While monopoles have the advantage that their charge would not be quickly neutralized,
their realism is questioned by never having been observed. However, they are thought to be formed
in the early universe via symmetry breaking in a wide range of field theories.

In this paper, we study the merger of two black holes initially in quasi-circular orbits with
very weak charges that do not affect the metric. We analyze the gravitational and electromagnetic
wave production and comment on the model by Zhang~\cite{Zhang:2016rli}.

\section{Numerical Setup}
\label{sec:numerical_setup}

To model the merger of two charged black holes, we use our long standing code that incorporates fully nonlinear general relativity coupled to an electromagnetic field.
However, even within this framework, the model is not complete until the particular environment outside the black holes is specified. If the magnetosphere is filled with a tenuous plasma, one could
assume that magnetic forces dominate over inertial forces, so that currents are generated to cancel the Lorentz forces (i.e., the force-free approximation).
Such a condition cannot hold near the electrically charged black holes however, and instead
perhaps the ``charge starved'' environment (see Ref.~\cite{25354}) resembling electrovacuum is more appropriate here. 
And so we consider first electrically charged black holes in electrovacuum (or equivalently black holes with
magnetic charges), and then consider magnetically charged black holes within a force-free environment in the subsequent
section.

Therefore, we solve the Einstein-Maxwell equations to model the electromagnetic field and the strong gravity dynamics in the current scenario of binary charged black holes. We use the BSSN formulation \cite{1995PhRvD..52.5428S, 1999PhRvD..59b4007B} to implement the Einstein equations, and  to implement the Maxwell equations we use the formulation described in Refs.~\cite{Komissarov:2004ms,2010PhRvD..81h4007P}. We have employed this formulation to study other electrovacuum scenarios involving black holes~\cite{Palenzuela:2009yr,2010PhRvD..81h4007P,2010PhRvD..81f4017M}.

We consider a binary of charged black holes
described by individual masses $m_i$ 
and charges $Q_i$. It is convenient
to use a non-dimensional quantity for 
the charge $q_i=Q_i/m_i$. 
In this first work
we restrict ourselves to non-spinning cases ($a_i=0$)
and small values of $q$, so that the energy associated
with the electromagnetic field remains several orders of 
magnitude smaller than that of the gravitational field.
The electromagnetic field therefore
has a negligible influence on the dynamics of the black holes,
and its feedback on the spacetime is not included in our simulations.
Notice that, for small values of $q$,  the electromagnetic luminosity should scale with $q^2$.
The force-free system, as it involves magnetic reconnection and current
sheets, is nonlinear and so this scaling is not exact. Nevertheless,
the luminosities and energies are sufficiently coarse-grained that we find the scaling holds quite well.
For the cases in which both black holes are charged, we study only those with equal charge magnitudes.

We adopt initial data corresponding to quasi-equilibrium, equal-mass, non-spinning black holes with $q=0.01$ initially separated by a distance of $\approx 8 M$.
With this separation, the merger takes place after about 4-5 orbits. The initial data for the metric
is constructed by superposing two boosted, uncharged black holes. The electromagnetic
field is initialized as a superposition of the fields of a boosted isolated charge for each BH.
Notice that while the electromagnetic constraints are satisfied by the initial data,
the Hamiltonian and momentum constraints are only approximately satisfied by this 
construction. 
To explore the large charge regime, one would need to solve the full initial data problem by solving the full set of elliptic constraint equations.

To extract physical information, we monitor the 
Newman-Penrose radiative scalars; in particular, the electromagnetic ($\Phi_2$) and gravitational ($\Psi_4$) radiative
scalars. These scalars are computed by contracting the Maxwell and the Weyl tensors respectively with a suitably defined null tetrad 
\begin{eqnarray}
  \Phi_2 = F_{ab} n^a \bar m^b ~~,~~ \Psi_4 = C_{abcd} n^a \bar m^b n^c \bar m^d ~~,
\end{eqnarray}
and they account for the energy carried off by outgoing waves at infinity.
The total energy flux (luminosity) in both electromagnetic and gravitational waves are
\begin{eqnarray}
  L_\mathrm{EM} &=&  \lim_{r \rightarrow \infty}  \int \frac{r^2}{2 \pi} |\phi_2|^2 d\Omega ~,
\label{FEM} \\
  L_\mathrm{GW} &=&  \lim_{r \rightarrow \infty} \int \frac{r^2}{16 \pi} \left| \int_{\infty}^t \Psi_4 dt' \right|^2 d\Omega~. 
\label{FGW}
\end{eqnarray}
In order to estimate the collimation as a function  of time, we compute the luminosity, $L_\theta$, within a jet opening angle $\theta$ with respect to the perpendicular to the orbital plane. This luminosity is normalized with respect to the luminosity over the hemisphere, $L_{90^\circ}$, so that for perfectly collimated emission (i.e. within the opening angle $\theta$) this normalized quantity, $L_\theta/L_{90^\circ}$, is just unity.

We adopt finite difference techniques on a regular Cartesian grid to solve the overall system numerically. To ensure sufficient resolution in an efficient manner we employ adaptive mesh refinement~(AMR) via the \had\ computational infrastructure that provides distributed, Berger-Oliger
style AMR~\cite{had_webpage,Liebling} with full sub-cycling
in time, together with an improved treatment of artificial boundaries~\cite{Lehner:2005vc}. A fourth order accurate spatial discretization satisfying a summation by parts rule together with a third order accurate in time Runge-Kutta integration scheme are used to help
ensure stability of the numerical implementation~\cite{Anderson:2007kz}. We adopt a Courant parameter of $\lambda = 0.25$ so that $\Delta t_l = 0.25 \Delta x_l$ on each refinement level $l$. On each
level, one has full sub-cycling in time and therefore one ensures that the Courant-Friedrichs-Levy~(CFL) condition dictated by the principal part of
the equations is satisfied. 
   Our evolutions generally adopt seven levels of refinement with a
   finest resolution of $\Delta x^i = 0.04 M$. We use a self-shadow
   method as our refinement criterion, although the truncation error
   is dominated mainly by the spacetime quantities.
Because, as mentioned, this code has been used extensively for a number of other projects, it has already been rigorously tested.
We also test for convergence, charge conservation, and divergencelessness of the magnetic field (outside the black hole horizons) here.
We use geometric units in which $G=c=1$ unless otherwise stated.

\section{Results}

\begin{figure}[h]
\centering
\includegraphics[width=8.2cm,angle=0]{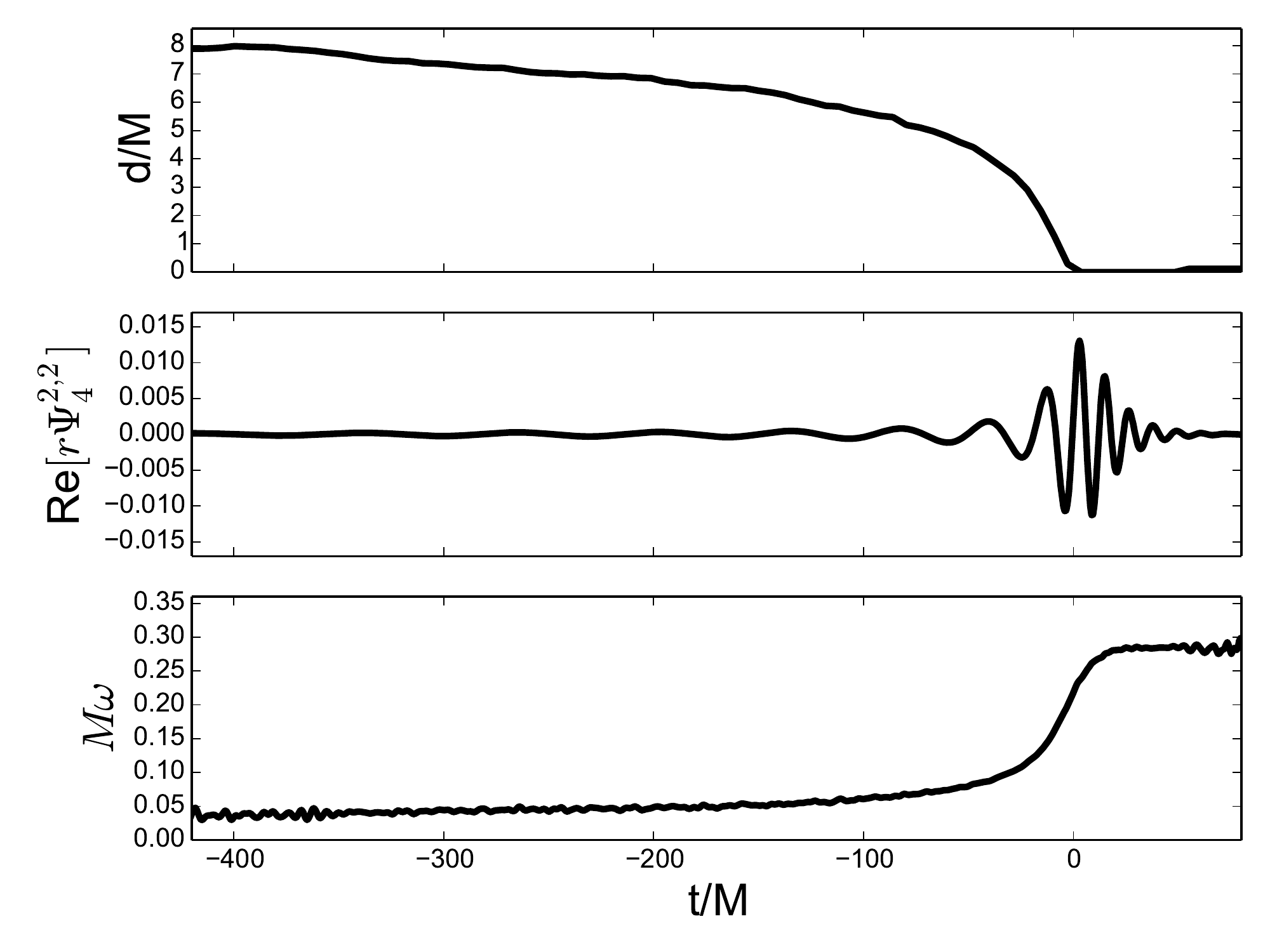}
\caption{ The dynamics of the binary system and its gravitational radiation. \textbf{Top:} The 
separation, $d$, as a function of time, obtained by the coordinate distance between the two minima of the conformal factor $\chi$. \textbf{Middle:} The real part of the $l=m=2$ mode of the Newman-Penrose scalar $\Psi_4$. \textbf{Bottom:} The angular frequency of the binary obtained from the main gravitational-wave mode.
} 
\label{fig:trajectories}
\end{figure}

Because we study the binary in the weakly charged limit, the dynamics of the binary
are independent of initial charge configuration.
The binary orbits for 4-5 cycles before merging into
a single spinning black hole as shown
in Fig.~\ref{fig:trajectories}.
We have labeled the times such that $t=0$ occurs when the gravitational wave luminosity peaks (as shown
in Fig.~\ref{fig:luminosities}).

A simple calculation
indicates that the gravitational wave radiation, at the lowest order, has a quadrupolar structure~\cite{1963PhRv..131..435P}. For equal mass binaries, the luminosity carried away by these waves goes as $L_\mathrm{GW} \propto (M \omega)^{10/3}$,
where we have used the Keplerian relation $\omega^2 r^3 = M$.

\subsection{Electrovacuum environment}

\begin{figure}[h]
\centering
\includegraphics[width=8.2cm,angle=0]{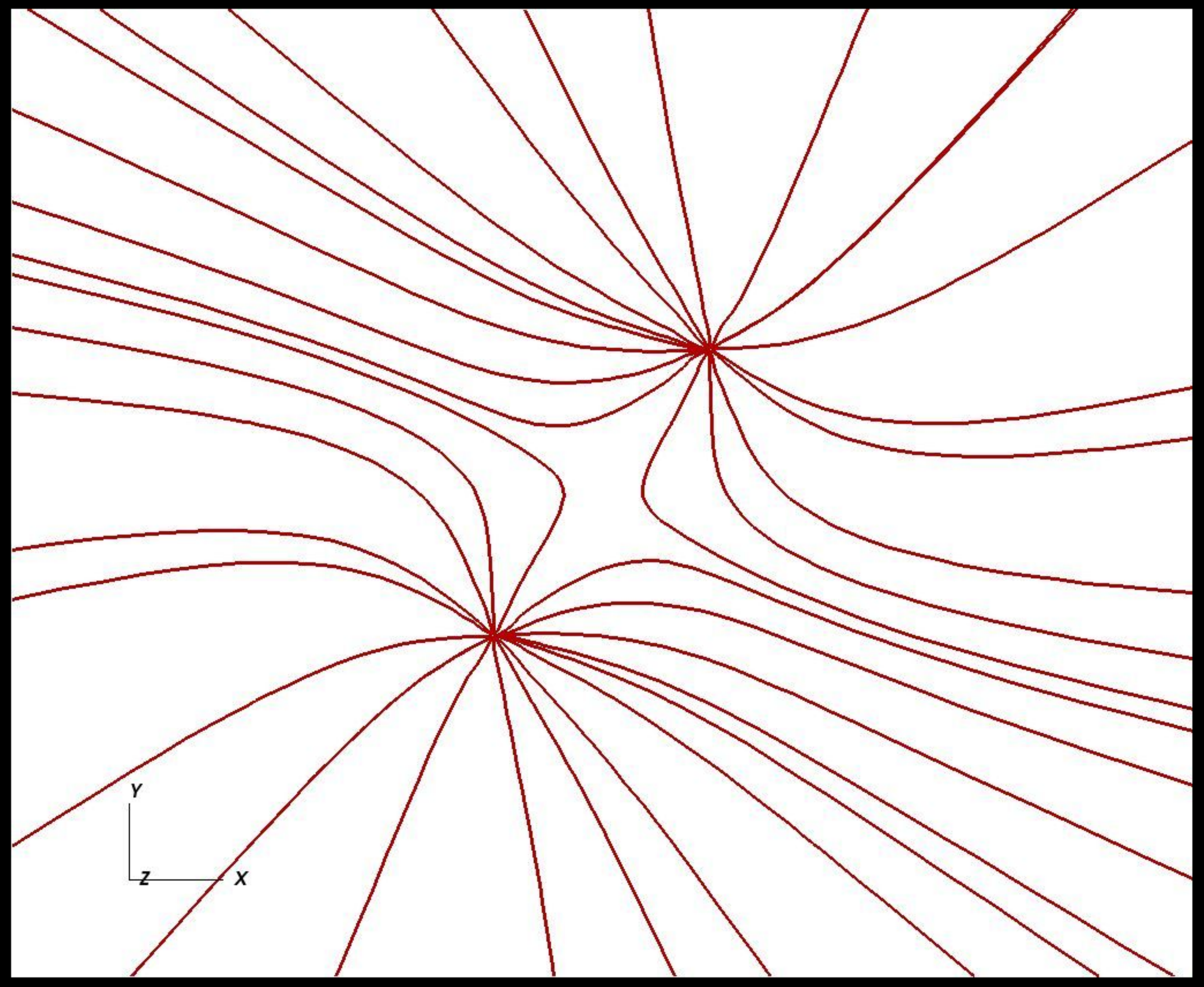} \\
\vspace{0.1in}
\includegraphics[width=8.2cm,angle=0]{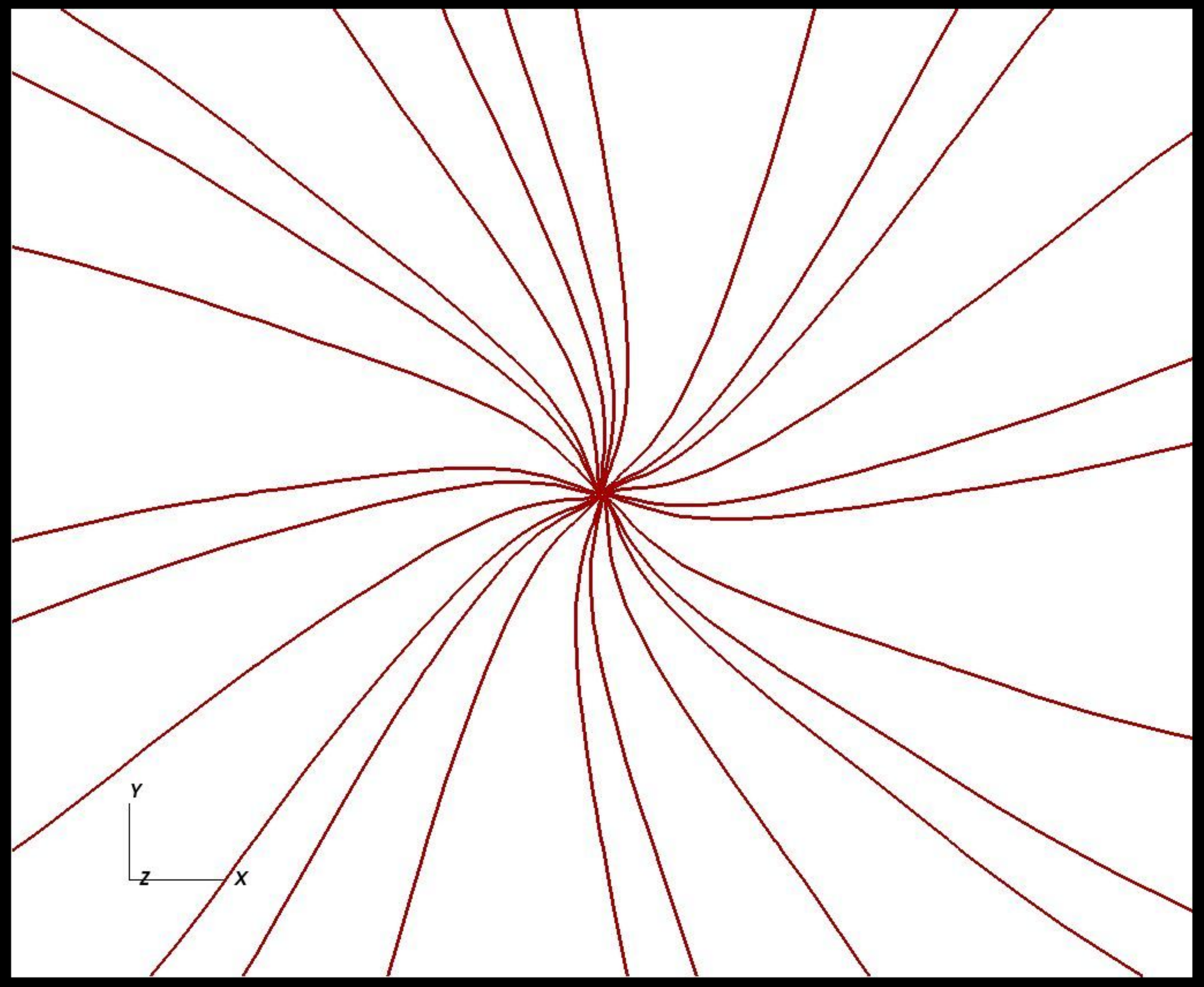} 
\caption{The electric field in the equatorial plane, $z=0$, for the \textit{electrovacuum case}.
Only a centered region of the computational domain is shown, spanning
a width of four times the initial separation.
\textbf{Top:} Before and \textbf{Bottom:} after the merger of the BBH system with $\{+,+\}$ charge configuration. 
After the merger, the electric field quickly settles to that of a spinning charge.
The black holes are located where the field lines converge.
} 
\label{fig:Efields}
\end{figure}

We consider three different charge 
configurations $\{+,+\}$, $\{+,0\}$ and $\{+,-\}$ and focus on
the electromagnetic luminosity of each case.
Notice that although in this section we consider black holes with electric charge,
these results apply to magnetically
charged black holes as well.
The $\{+,+\}$ binary consists of two positively charged black holes that ultimately merge into  a single, charged, spinning (Kerr-Newman) black hole (see the
electric field configuration in Fig.~\ref{fig:Efields}). The $\{+,-\}$ binary consists of black holes with
opposite charge and thus no net charge. 
The $\{+,0\}$ binary has one uncharged black hole and represents a superposition of the other
two configurations.

The electromagnetic luminosity is generally dominated by
the dipolar radiation so that for a system of two orbiting charges the luminosity scales as
$L_\mathrm{EM}^\mathrm{dip} \propto  q^2 (M \omega)^{8/3}$. 
If no dipole is present, the next dominant term for an orbiting pair of charges is the quadrupolar contribution $L_\mathrm{EM}^\mathrm{quad} \propto q^2 (M \omega)^{10/3}$. 

\begin{figure}[h]
\centering
\includegraphics[width=8.2cm,angle=0]{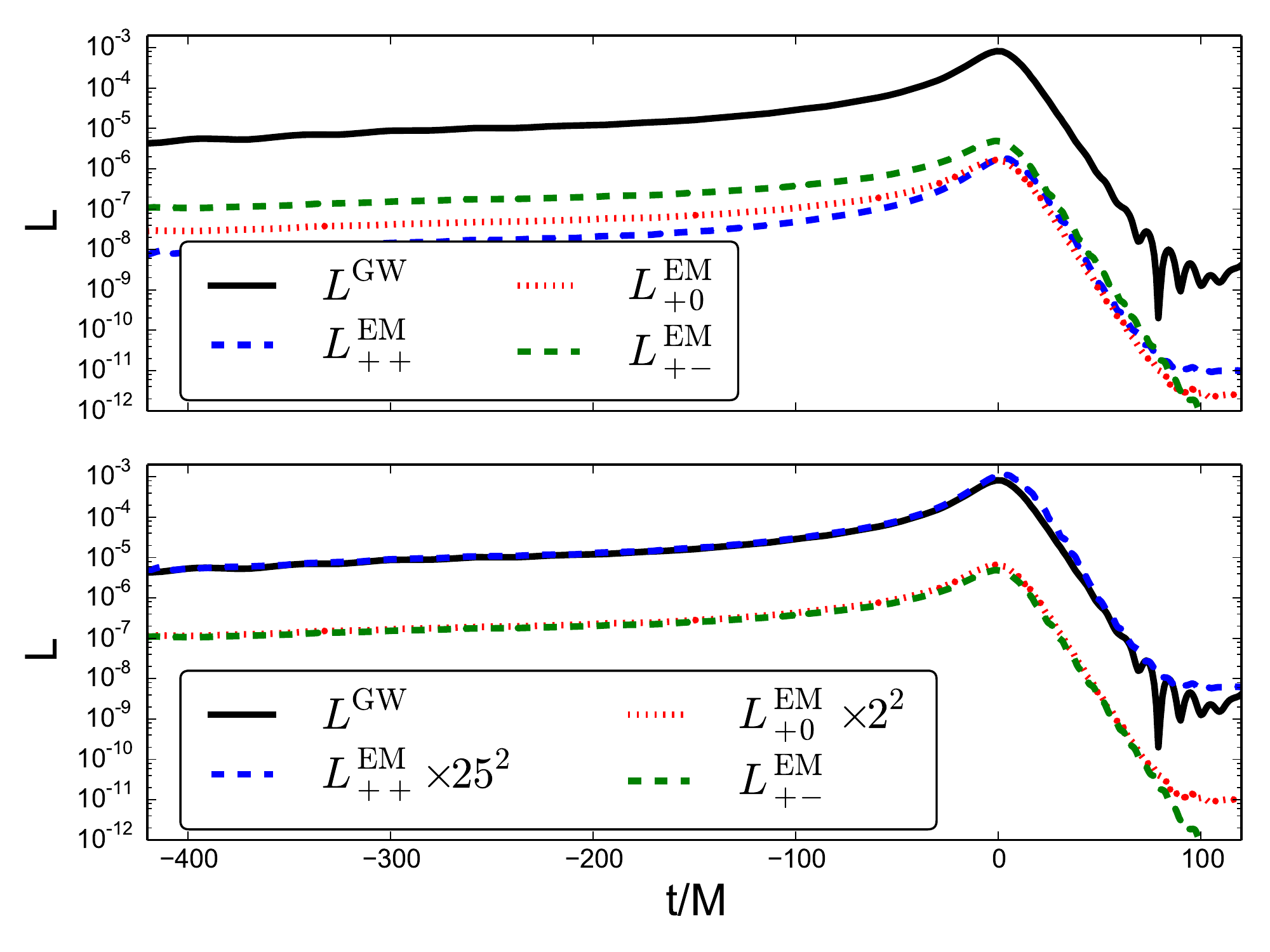}
\caption{ Electromagnetic and gravitational luminosities for the different \textit{electrovacuum configurations}. 
\textbf{Top:}
The gravitational luminosity (solid black) is the same among all cases.
The electromagnetic luminosities vary 
for the different charge configurations.
\textbf{Bottom:}
The $\{+,+\}$ EM luminosity has been rescaled to show that the electromagnetic luminosity is
proportional to the gravitational luminosity.
Likewise, the $\{+,0\}$ EM luminosity has been rescaled showing that it is roughly a quarter 
of the luminosity of the $\{+,-\}$ case.
} 
\label{fig:luminosities}
\end{figure}

\begin{figure}[h]
\centering
\includegraphics[width=8.8cm,angle=0]{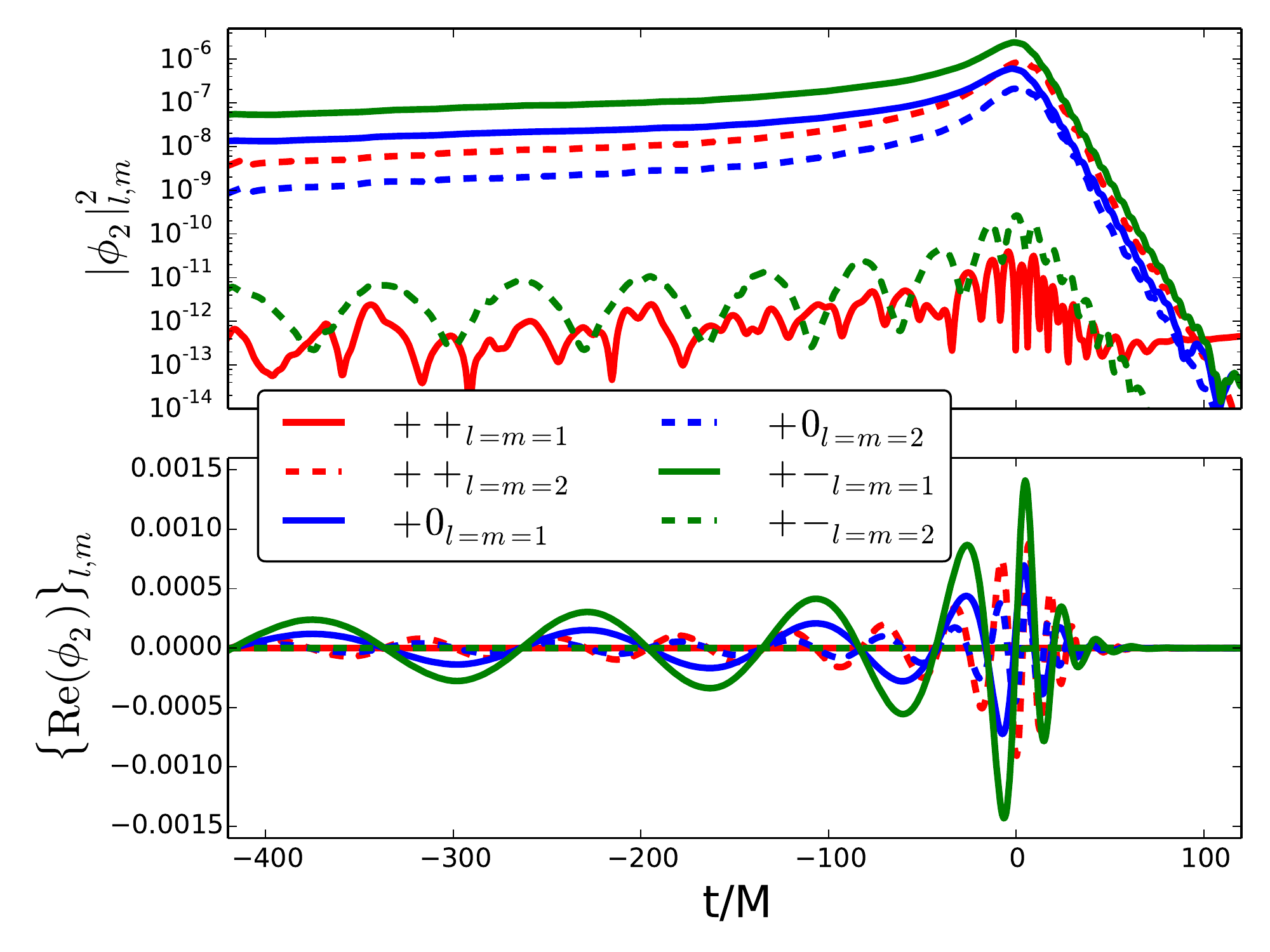} 
\caption{The main spin-weighted spherical-harmonic modes of the Newman-Penrose scalar $\Phi_2$ describing the electromagnetic radiation for the
different \textit{ electrovacuum configurations}. \textbf{Top:} The norms of the main modes. \textbf{Bottom:} The real part of the main modes.
} 
\label{fig:modes}
\end{figure}

We display the
electromagnetic luminosity for each case in Fig.~\ref{fig:luminosities}.
The most radiative case is $\{+,-\}$, followed by the
$\{+,0\}$ case, and finally the $\{+,+\}$. 
The high luminosity of the $\{+,-\}$ case can be understood in terms of its dipole
moment. 
The most radiative electromagnetic radiative mode arises from the acceleration of the dipole moment, which will produce dipolar radiation corresponding to the $l=m=1$ mode. 
The  $\{+,+\}$ case has no dipole moment by symmetry, and the $\{+,0\}$  case has half the dipole
moment of the $\{+,-\}$ case.
Instead, the  $\{+,+\}$ binary radiates quadrupolar radiation giving it the same frequency
dependence as the gravitational wave radiation.
The case 
$\{+,0\}$ has both dipolar and quadrupolar contributions, although the dipolar is dominant. 
The differences among the modes for the different configurations are shown in Fig.~\ref{fig:modes}. 
The $\{+,+\}$ case emits radiation mainly near the equatorial plane while the $\{+,-\}$ case radiates more isotropically.

\begin{figure}[h]
\centering
\includegraphics[width=8.2cm,angle=0]{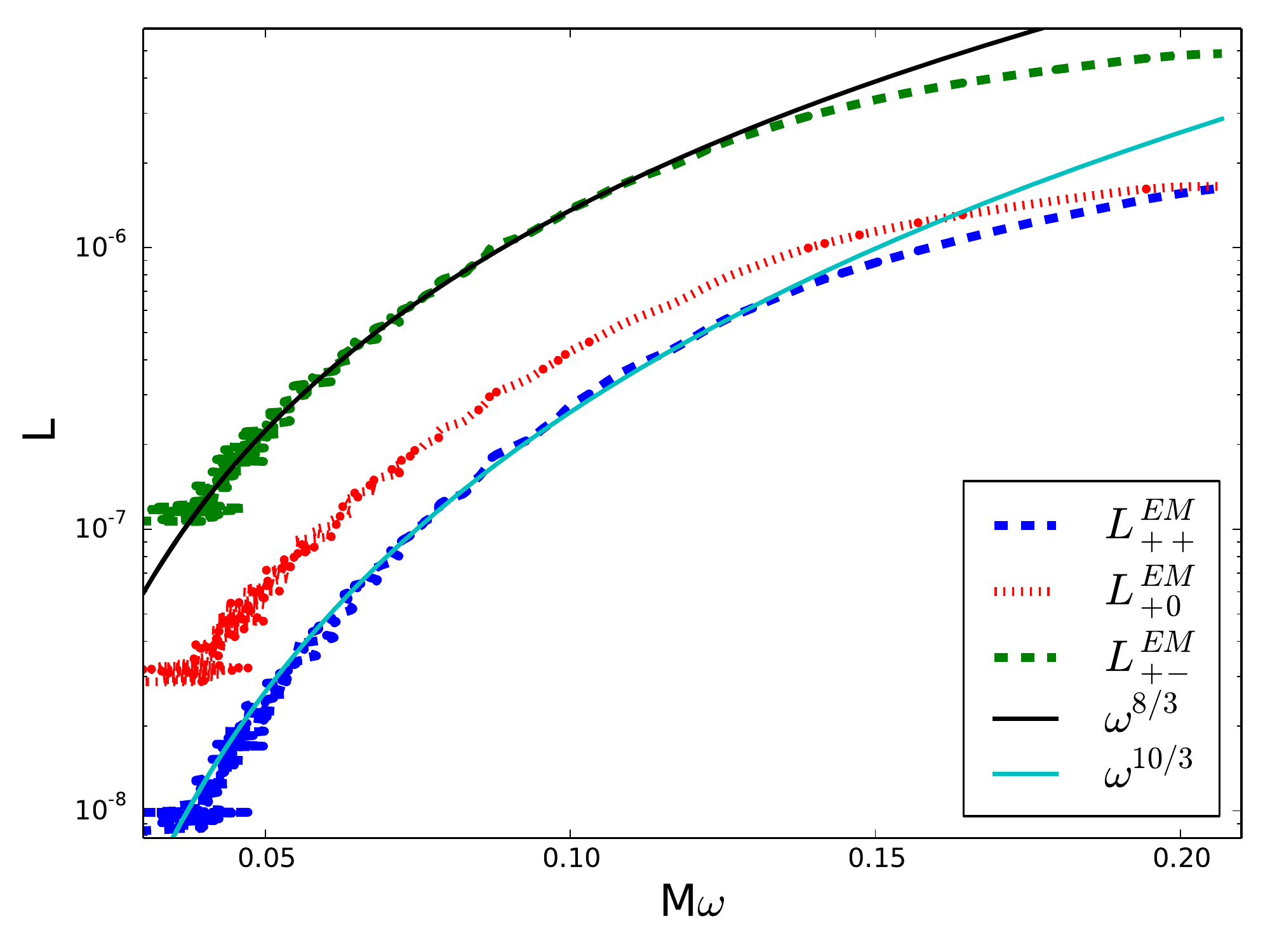}
\caption{ Electromagnetic luminosities 
as a function of the angular frequency of the binary for the \textit{electrovacuum cases}. 
Fits of these luminosities to a power law in the frequency are also shown (solid curves). 
Notice that the dependence of the luminosities 
on the orbital frequency follows that expected from a pair of orbiting charges, at least up to the plunge.
} 
\label{fig:freq_luminosities}
\end{figure}

We display the luminosities in terms of the angular frequency of the binary in
Fig.~\ref{fig:freq_luminosities}.
We also fit these luminosities at early
times before the last part of the coalescence
of form $a \omega^b$ for real constants $a$ and $b$,
 obtaining
\begin{eqnarray}
   L_\mathrm{++} &=& 1.3 \times 10^{-8} \left[ \frac{q}{0.01}\right]^{2}
   \left[ \frac{M \omega}{0.04}\right]^{10/3} \\
   L_\mathrm{+0} &=& 3.8 \times 10^{-8} \left[ \frac{q}{0.01}\right]^{2}
   \left[ \frac{M \omega}{0.04}\right]^{8/3} \\
   L_\mathrm{+-} &=& 1.3 \times 10^{-7} \left[ \frac{q}{0.01}\right]^{2}
      \left[ \frac{M \omega}{0.04}\right]^{8/3} 
\end{eqnarray}
in good agreement with the expected analytic estimates. Notice
however that the luminosities deviate from the fits at frequencies
higher than $M \omega=0.12$ (i.e., separations smaller than $d=4M$),
when the system departs from a slow inspiral and relativistic effects becomes more predominant.

Integration of the luminosities in time gives estimates of the 
total radiated energies
\begin{equation}
   E_\mathrm{GW}/M = 3 \times 10^{-2}  ,~~
   E_\mathrm{EM}/M \approx 10^{-4} \left[ \frac{q}{0.01}\right]^{2}
\end{equation}
where the coefficients for the electromagnetic energies  range over $\{7.6\times10^{-5}, 8.6\times10^{-5},2.7\times10^{-4}\}$ for the $\{++,+0,+-\}$
configurations.

\subsection{Force-free environment}

We consider now magnetically charged black holes immersed
in a magnetically dominated, low density plasma such that the
Lorentz force vanishes, leading to the force-free approximation~\cite{Goldreich:1969sb,1977MNRAS.179..433B}.
Such a regime excludes
electrically charged black holes, and so we will only consider the binaries with a magnetic monopole charge.
 
\begin{figure}[h]
\centering
\includegraphics[width=8.2cm,angle=0]{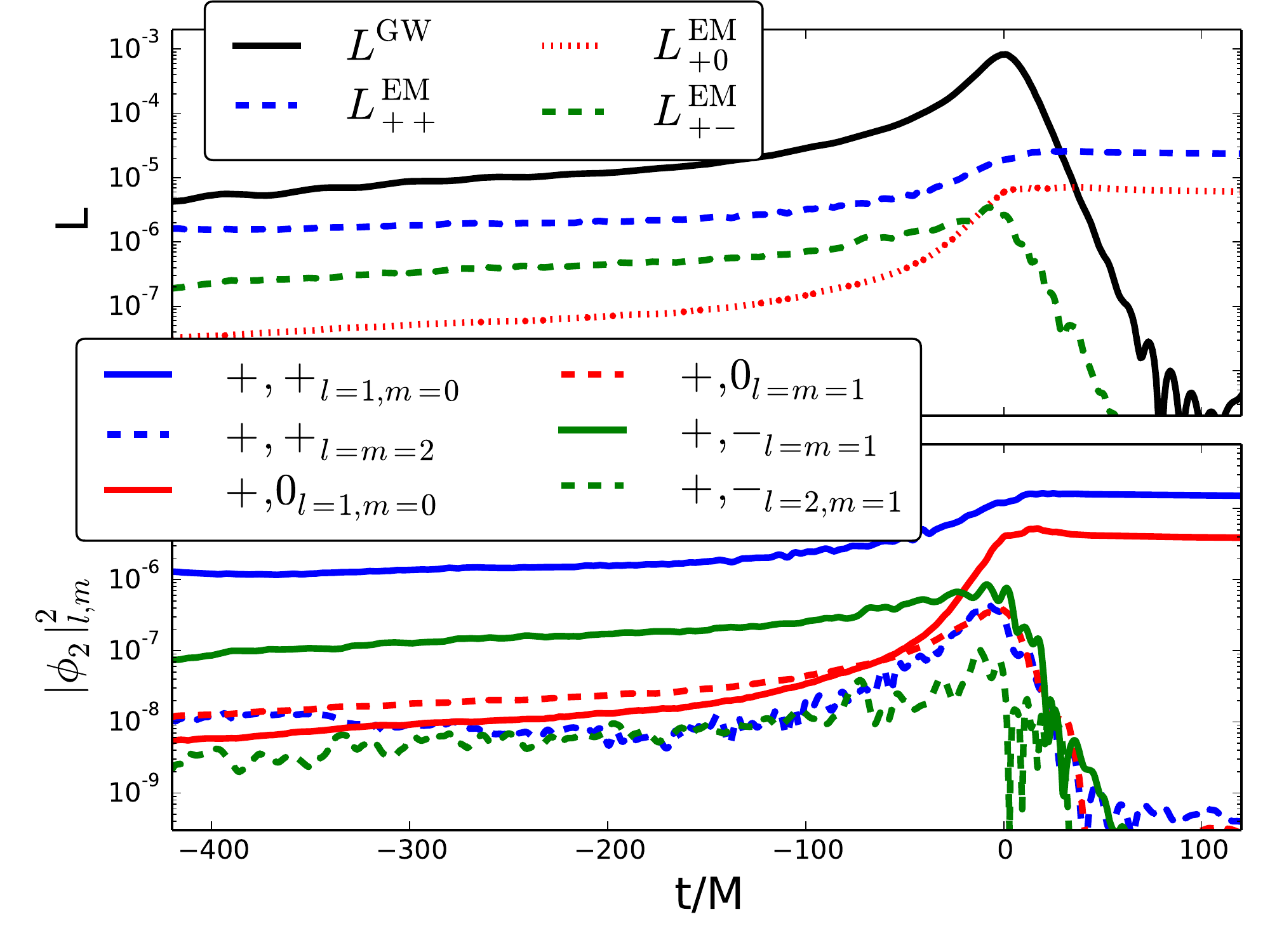}
\caption{ Electromagnetic and gravitational luminosities from the different cases in a \textit{force-free environment}. 
\textbf{Top:}
The gravitational luminosity (solid black) is shown for comparison with the previous electrovacuum cases.
The electromagnetic luminosities vary 
for the different charge configurations.
\textbf{Bottom:}
The norms of the main spin-weighted spherical-harmonic modes of the Newman-Penrose scalar $\Phi_2$, showing that the radiation
is mainly dipolar. 
} 
\label{fig:luminosities_ff}
\end{figure}

We evolve the same configurations as in the electrovacuum case and display the 
electromagnetic luminosities in
Fig.~\ref{fig:luminosities_ff}. Once again, the luminosity is still dominated generally by the dipolar radiation.
The most radiative case is now the $\{+,+\}$, followed by the
$\{+,0\}$ case, and finally the $\{+,-\}$.  
Note that in contrast to the electrovacuum case, the force-free evolutions
tend to be less smooth having to resolve magnetic reconnection and current sheets, particular with the $\{+,-\}$ case.

\begin{figure}[h]
\centering
\includegraphics[width=8.2cm,angle=0]{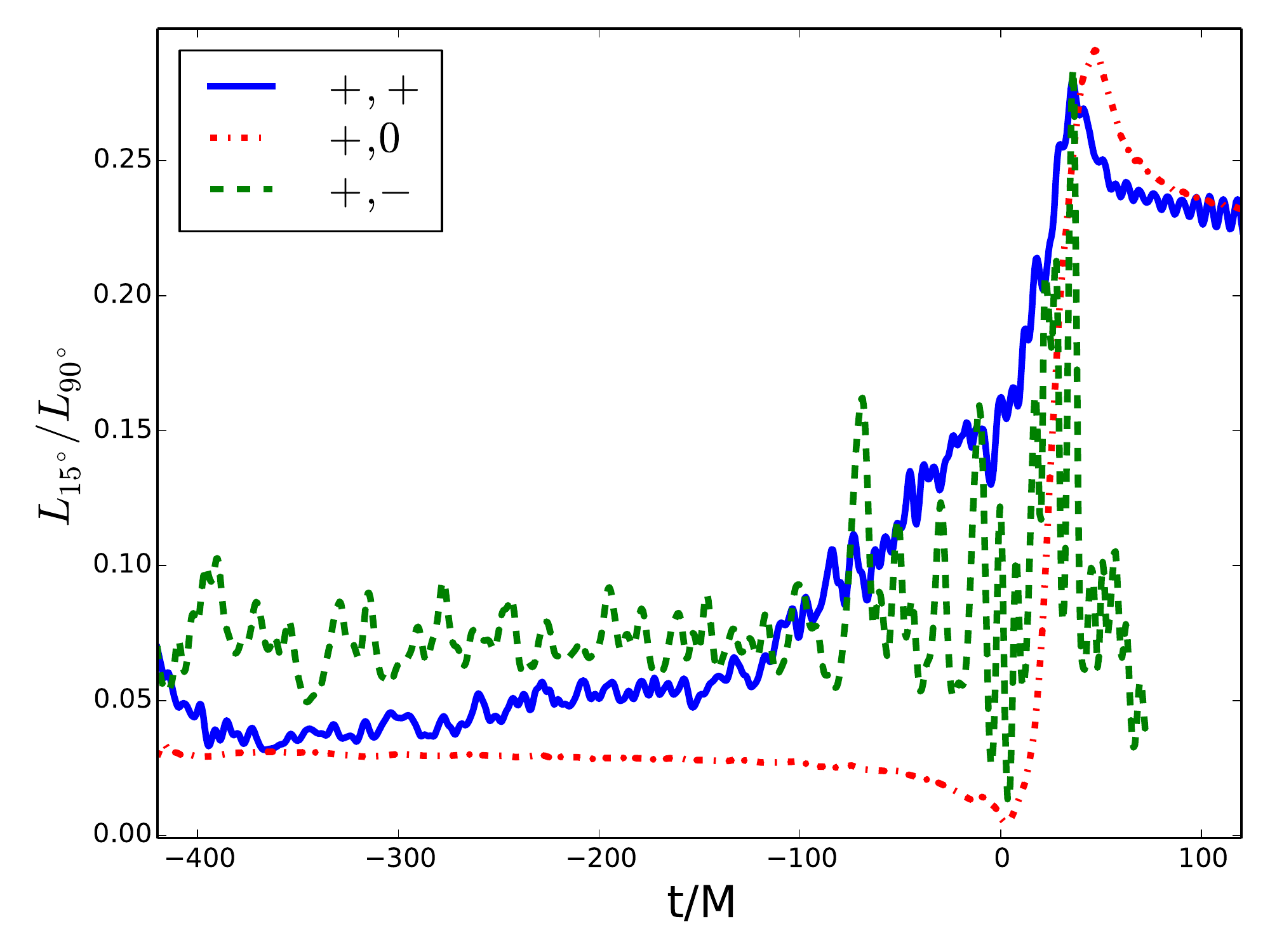}
\caption{
The normalized luminosities emitted in a jet opening angle of $15^\circ$ (which estimates the collimation of the radiation) 
for the \textit{force-free} cases.
The $\{+,-\}$ case is quite collimated during the inspiral and merger. In contrast,  the collimation of the other two cases 
increases towards the merger as would be expected from a spinning black hole with magnetic monopole charge through the Blandford-Znajek mechanism.
} 
\label{fig:collimation}
\end{figure}

In addition, the luminosity of the $\{+,-\}$ case is quite collimated
during the inspiral and merger (the collimations are displayed in Fig.~\ref{fig:collimation}).
In contrast, the $\{+,+\}$ luminosity is not
as collimated during the inspiral, but its collimation increases towards the merger.
The post-merger collimation is consistent with the  Blandford-Znajek emission from a spinning
black hole with a magnetic monopole. 
The collimation of the $\{+,0\}$ case is the smallest during the inspiral 
and increases only during the merger  when the remnant spinning black hole is formed, again consistent
with the emission through the Blandford-Znajek mechanism.

This behavior,
completely opposite to the electrovacuum case, can be understood
by analyzing the mode decomposition of $\phi_2$ (shown in the bottom panel of Fig.~\ref{fig:luminosities_ff}).
The strongest electromagnetic mode in the $\{+,+\}$ configuration corresponds to the $l=1,m=0$, followed by the $l=m=2$. 
The magnetic field structure of the late inspiral stage reveals that the binary resembles a single spinning black hole
in a force-free environment. Previous studies of spinning black holes have shown that for fairly generic magnetic field topologies, such a 
configuration radiates axisymmetrically with a $m=0$ mode~\cite{1977MNRAS.179..433B}. That the binary is not axisymmetric
means that there is an additional, sub-dominant component, namely the $l=m=2$ mode.

Another way of interpreting the results during the inspiral
phase is by considering the motion of one of the black holes.
This motion, relative to the other BH, will produce deformations of the 
magnetic field, and, in a force-free environment, these perturbations 
will propagate along the magnetic field lines as \Alfven waves, carrying away energy from the system. These waves will be stronger where the magnetic 
field lines and the black hole velocity are perpendicular to each other. In particular, the radiation will be more efficient along the polar caps where the magnetic fields are not strongly affected by the interaction with the other charged black hole.
The high degree of collimation of the
radiation in the $\{+,+\}$ case supports this interpretation. 

The  $\{+,-\}$ case is very different, since the magnetic field
lines from one black hole reconnect with the magnetic field lines from the other one, 
forming a dipole rotating in the equatorial plane. Perturbations of the magnetic field travel
from one black hole to the other, but only the open magnetic field lines can transport 
outgoing radiation. Thus, the luminosity---which is dominated by the $l=m=1$ mode---is reduced compared to the $\{+,+\}$ configuration, and tends to zero after the merger when the charges neutralize. 

The $\{+,0\}$ case is not a clean superposition of the other two cases as appeared in the
electrovacuum case because the force-free equations are nonlinear.
The single most luminous modes of the other two force-free cases appear as the two most significant modes for
this case, namely the dipolar radiation mode ($l=m=1$) and the axisymmetric $l=1, m=0$ mode. 
One can understand this by considering the 
uncharged black hole as it moves through the magnetic field sourced by the charged BH. 
As in the electrovacuum case, its motion induces
an electric charge separation on the uncharged BH horizon, but, unlike the electrovacuum case,
the force free environment dictates the interaction of the fields of the two black holes.
Interestingly, the total luminosity prior to merger is quite similar to that of 
the corresponding electrovacuum case.

Note that the three configurations differ significantly in how luminous they are as well as
how collimated they are because of the interaction between the black holes. If somehow future electromagnetic counterparts lend support to this 
scenario, perhaps the charge configuration of the binary can be extracted from these features.

The disparate behavior of the post-merger luminosities (top panel of Fig.~\ref{fig:luminosities_ff}) are straight forward 
to explain. In particular, both the $\{+,+\}$ and the $\{+,0\}$ cases produce spinning BHs
 with magnetic charge in a force-free environment.
This solution should correspond precisely to the well-known stationary, radiating solution found by Blandford and Znajek with a monopole
magnetic field~\cite{1977MNRAS.179..433B}.
That the  $l=1, m=0$ mode
plateaus while the other modes quickly turn off is consistent with 
the post-merger solution approaching the Blandford-Znajek solution (see the bottom panel of Fig.~\ref{fig:luminosities_ff}).
In contrast, the $\{+,-\}$ produces an uncharged black hole and so the luminosity
simply shuts off.

\begin{figure}[h]
\centering
\includegraphics[width=8.2cm,angle=0]{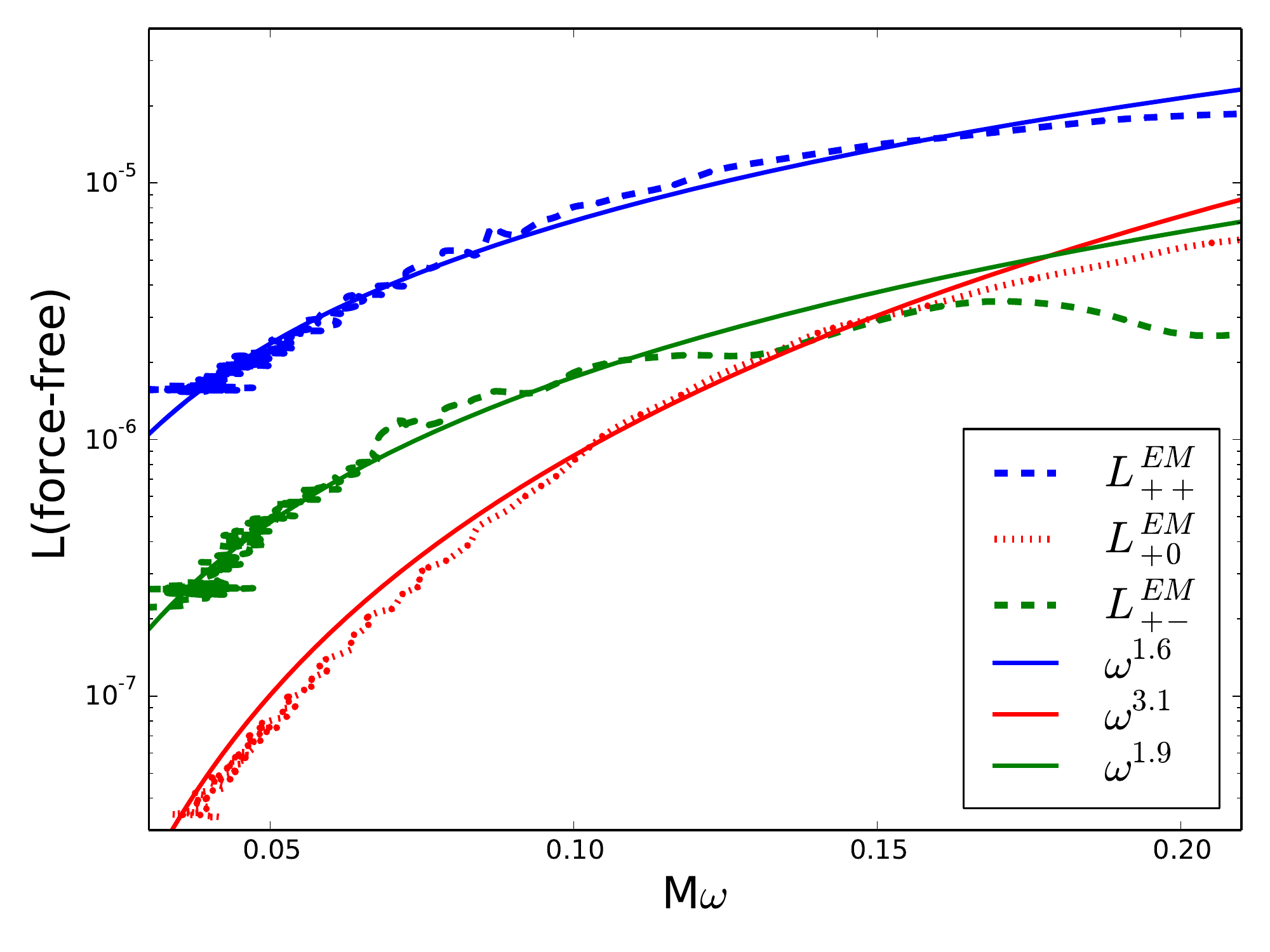}
\caption{ Electromagnetic luminosities of the \textit{force-free cases}
as a function of the angular frequency of the binary. 
Fits of these luminosities to a power law in the frequency are also shown (solid curves).
} 
\label{fig:freq_luminosities_ff}
\end{figure}

We display the luminosities in terms of the angular frequency of the binary in
Fig.~\ref{fig:freq_luminosities_ff}.
We again fit these luminosities at early
times before the last part of the coalescence,
obtaining
\begin{eqnarray}
   L_\mathrm{++} &=& 1.7 \times 10^{-6} \left[ \frac{q}{0.01}\right]^{2}
   \left[ \frac{M \omega}{0.04}\right]^{1.6} \\
   L_\mathrm{+0} &=& 4 \times 10^{-8} \left[ \frac{q}{0.01}\right]^{2}
   \left[ \frac{M \omega}{0.04}\right]^{3.1} \\
   L_\mathrm{+-} &=& 3 \times 10^{-7} \left[ \frac{q}{0.01}\right]^{2}
      \left[ \frac{M \omega}{0.04}\right]^{1.9} ~.
\end{eqnarray}
Although there are no analytic estimates to compare with these results, it is reasonable to consider a similar force-free scenario given by a binary black hole immersed in an external magnetic field. Numerical simulations for such a 
system yield a dependence $L \propto \omega^{5/3-6/3}$ \cite{2012ApJ...749L..32M},
which is a range consistent with our results, albeit with a different magnetic field topology. 
 
Finally, integration of the luminosities in time gives estimates of the total radiated energies an order of magnitude larger than
the electrovacuum case, namely
\begin{equation}
   E_\mathrm{EM}/M \approx 10^{-3}  \left[ \frac{q}{0.01}\right]^{2}
\end{equation}
where the coefficients for the electromagnetic energies  range over $\{5\times10^{-4}, 6 \times10^{-4},5\times10^{-3}\}$ for the $\{++,+0,+-\}$
configurations.

\section{Discussion}

Because our primary interest in this study is to constrain the presumed charge of
the binary detected by GW150914, we introduce a mass of $M \approx 65 M_{\odot}$ and find
an
estimate of the peak
electromagnetic luminosity and total energy during the coalescence of
\begin{eqnarray}
   L^\mathrm{peak}_\mathrm{EM} &\approx& 10^{53} \mathrm{ergs/s}
    \left[ \frac{q}{0.01}\right]^{2} \\
   E_\mathrm{EM} &\approx& 10^{50} \mathrm{ergs} 
    \left[ \frac{q}{0.01}\right]^{2}
    \left[ \frac{M}{M_{\odot}}\right].
\end{eqnarray}
Therefore, if the Fermi~GBM detection of $10^{49}$\,ergs/s is indeed coincident with
GW150914, then our results suggest that $q \approx 10^{-4}$
if
the reason the binary was observable as a weak sGRB is because of its charge. These
results are mostly consistent with the estimates of Zhang~\cite{Zhang:2016rli}. It should
also be noted that one generally expects that the electromagnetic energy we measure here
will couple to the environment surrounding the merger, ultimately producing the photons
seen on Earth. This process will introduce a delay between the gravitational and
electromagnetic signals as discussed by Zhang~\cite{Zhang:2016rli}. Here, we are
assuming the efficiency of this process is quite high.

For
such a binary to be the engine behind a (non-repeating) FRB with
luminosity of $10^{43}$\,ergs/s, the charge could be significantly
less, roughly $q\approx 10^{-10}$.
Of course, we are only considering energies and luminosities here because
the astrophysics of how this electromagnetic energy gets processed into
observable radiation is extremely difficult. And so we acknowledge that
hard X-rays and radio are very different bands and the details of
how such signals would be created would vary considerably.

As we acknowledge in the introduction, there is well deserved skepticism that a 
BBH could maintain its charge, even a charge of $q=10^{-4}$, until merger. Another possibility is that
one can switch the roles of electric and magnetic fields to consider a binary with at least one
magnetic monopole charge.
Magnetic monopoles are widely thought to be created in the
early universe, and cosmological models generally have to explain why no monopoles have been observed.
Inflation severely dilutes the density of monopoles while some models assert that primordial black holes might have accreted them~\cite{Stojkovic:2004hz}. 

Recent work suggests that a small window in black hole mass is allowed by various constraints for primordial black holes to be a primary source of dark matter and it happens that GW150914 fell in that 
window~\cite{Bird:2016dcv,Sasaki:2016jop,Clesse:2016vqa}. However, Ref.~\cite{Bird:2016dcv} does not consider the effects of any magnetic
field on the cosmological dynamics, and so the viability of a magnetically charged, primordial black hole binary forming, surviving, 
 and ultimately merging in the GW150914 event is by no means trivial.
Furthermore, such a binary would also have implications for the primordial magnetic field~\cite{Long:2015cza}.

Perhaps GW150914 contains at least one such monopole that accounts for the electromagnetic counterpart.
If so, then with $q=10^{-4}$, its surface magnetic field would be approximately $10^{14}$ G providing appropriate conditions
similar to those thought to power short gamma ray bursts from binary neutron star mergers. Our studies of the BBH within a
force-free environment showed even higher luminosities than the electrovacuum cases, although the $\{+,+\}$
may be excluded both because it may produce a long-lived post-merger signal (that was not observed) and because its
signal appears highly collimated and not likely to be observed in the first place.

Unlike the electrically charged case,
it would be difficult to neutralize any magnetic charge of the black holes. 
And because nothing rules out the existence of magnetic monopoles, perhaps
the strongest argument against such a scenario is simply that no one has seen one before. However, no one had seen a
$30 M_\odot$ black hole before, and so maybe with GW150914, we have now seen both.

Although no counterparts to GW151226, 
aLIGO's most recent detection~\cite{Abbott:2016nmj},
were found by Fermi\,GBM~\cite{Racusin:2016fko}, we look forward to an exciting era of both gravitational wave astronomy and its accompanying multi-messenger counterparts.

%
%
\vspace{0.5cm}

\begin{acknowledgments}
We thank Matthew Johnson for fruitful discussions, and Gaurav Khanna, Luis Lehner, and Patrick Motl for detailed comments on the
manuscript.
SLL gratefully acknowledges the assistance and hospitality of the
Universitat de les Illes Balears and the Spanish government where much
of this work was carried out.
This work was supported 
by the NSF under grants PHY-1308621~(LIU),
by NASA's ATP program through grant NNX13AH01G,
and
by Perimeter Institute for Theoretical Physics which is supported by the Government of Canada through Industry Canada and by the Province of Ontario through the Ministry of Economic Development \& Innovation.
CP acknowledges support from the Spanish Ministry of Education and
Science through a Ramon y Cajal grant and from the Spanish Ministry of
Economy and Competitiveness grant FPA2013-41042-P.
Computations were
performed at XSEDE and Scinet. 
\end{acknowledgments}

\bibliographystyle{utphys}
\bibliography{paper}

\end{document}